\documentclass[apj]{aastex62}
\usepackage{graphicx}	
\usepackage{amsmath}	
\usepackage{amssymb}	
\usepackage{natbib}

\newcommand\xray{\hbox{X-ray}}
\begin{document}

\title{Long-Timescale X-ray Variability of BAL and Mini-BAL Quasars}

\correspondingauthor{John Timlin}
\affiliation{Department of Astronomy \& Astrophysics, 525 Davey Lab, The Pennsylvania State University, University Park, PA 16802, USA }
\email{jxt811@psu.edu}

\author{John Timlin}
\affiliation{Department of Astronomy \& Astrophysics, 525 Davey Lab, The Pennsylvania State University, University Park, PA 16802, USA }

\author{W. N. Brandt}
\affiliation{Department of Astronomy \& Astrophysics, 525 Davey Lab, The Pennsylvania State University, University Park, PA 16802, USA }

\author{Shifu Zhu}
\affiliation{Department of Astronomy \& Astrophysics, 525 Davey Lab, The Pennsylvania State University, University Park, PA 16802, USA }


\begin{abstract}
We investigated the rest-frame $\approx0.1-5$ year \xray\ variability properties of an unbiased and uniformly selected sample of 24 BAL and 35 mini-BAL quasars, making it the largest representative sample used to investigate such variability. We find that the distributions of \xray\ variability amplitudes of these quasar populations are statistically similar to that of non-BAL, radio-quiet (typical) quasars.
\end{abstract}

\section{BAL X-ray Variability}

The key hallmark of broad absorption line (BAL) quasars is their high-velocity outflows originating from the quasar central regions. This outflowing gas often has sufficient column density to absorb the \xray\ emission from the quasar corona. For this reason, BAL quasars are often observed to be \xray\ weak compared to typical quasars (e.g., \citealt{Gibson2009}). Investigations of individual BAL quasars have also demonstrated that they can exhibit large-amplitude \xray\ variations, often likely due to changes of the absorbing gas (e.g., \citealt{Gallagher2004}). To investigate the long-term \xray\ variability of BAL quasars, \citet{Saez2012} compiled a (likely biased) sample of 11 BAL quasars that had multiple \xray\ observations split among different instruments. Using data from different instruments amplified the magnitude of the uncertainties in their flux measurements, and thus they found only three objects exhibited significant \xray\ variability. In this work, we present the \xray\ variability properties of a larger, representative sample of BAL and mini-BAL quasars and compare these results with the \xray\ variability properties of typical quasars.

We uniformly selected optically-bright ($i\leq20$) BAL quasars identified in the Sloan Digital Sky Survey (\citealt{York2000}) Data Release 16 quasar catalog \citep{Lyke2020} that had multiple \emph{Chandra} observations (mainly serendipitous). \xray\ image processing and source extraction for these quasars were performed in the same manner as in \citet{Timlin2020b} (hereafter, T20). We found 293 observations of 93 BAL and mini-BAL quasars that lie within the \emph{Chandra} footprint and have multiple sensitive observations.\footnote{See Section 3.3 in T20 for more details. This sample is presented here: \url{https://doi.org/10.5281/zenodo.4048833}} Count fluxes (cts cm$^{-2}$  s$^{-1}$) were computed for each observation, and variability amplitudes were computed between every epoch as the ratio of the earlier to the later observation. To ensure that no single quasar dominated the sample due to a large number of epoch permutations, the \xray\ light curve (LC) of each quasar was down-sampled to only three observations: the first and last epoch in the LC and a randomly drawn epoch in between (see Section~4 of T20 for more details).

\begin{figure}[h!]
\begin{center}
\includegraphics[scale=0.4,angle=0]{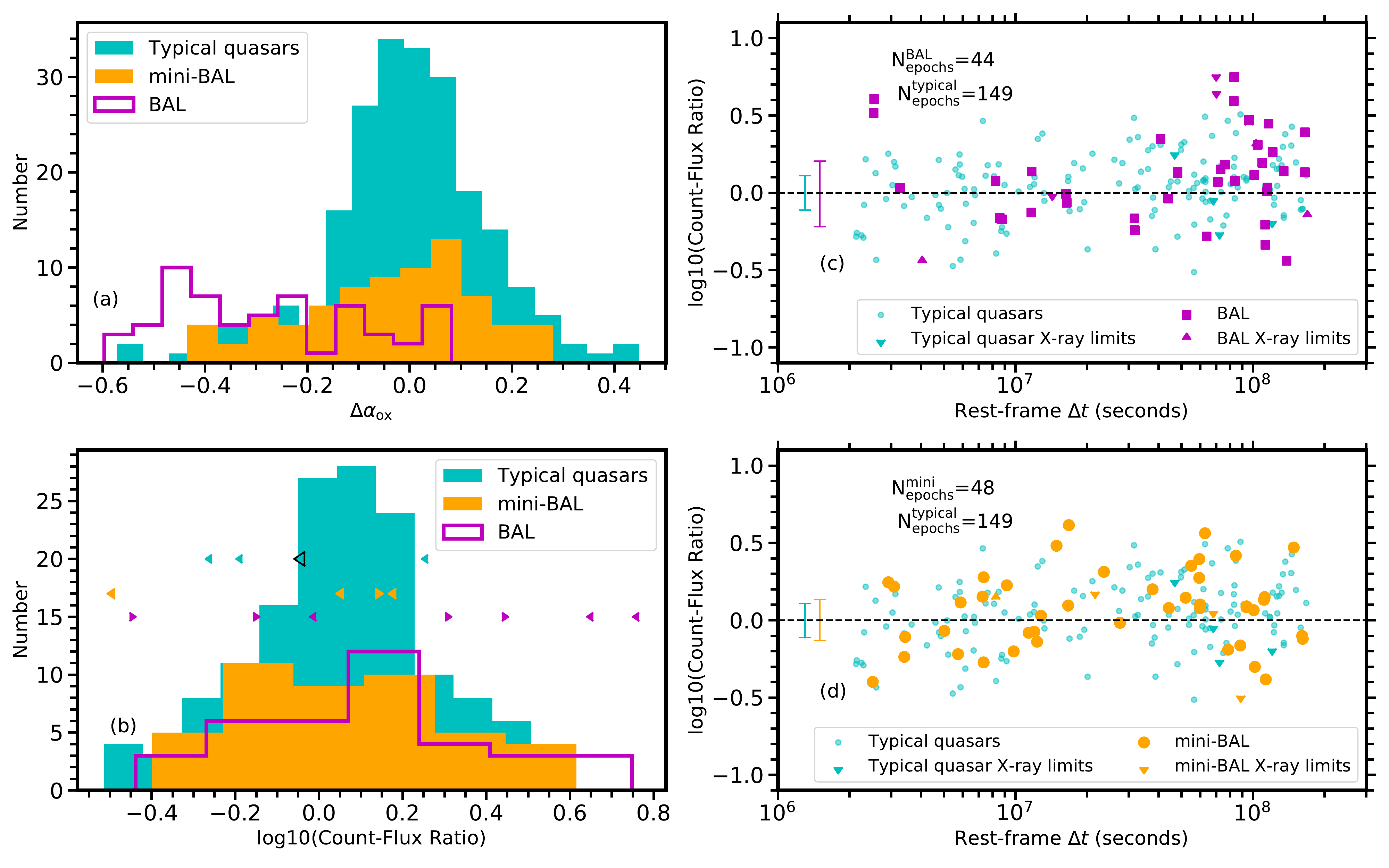}
\caption{Panel (a): The level of X-ray weakness, $\Delta\alpha_{\rm ox}$, compared to the expected value from Equation~3 of \citet{Pu2020}. As in, e.g.,\ \citet{Gibson2009} we find that the BAL quasars (open purple) are generally significantly X-ray weaker than the typical (filled cyan) and mini-BAL (filled orange) quasars, probably due to heavy absorption by the gas associated with the BAL outflow. Panel (b): The distribution of log(count-flux ratio) for the BAL, mini-BAL, and typical quasars. The arrows depict count-flux ratios in which one epoch is not X-ray detected, where the direction of the arrow represents whether the non-detection is in the numerator (left) or denominator (right). A Shapiro-Wilk test indicates that the BAL and mini-BAL quasar distributions are consistent with a Gaussian distribution, and we find that that the intrinsic dispersions of the \xray\ variability amplitudes for the three populations are statistically consistent. Panel (c): Count-flux ratio as a function of timescale, $\Delta t$, for BAL quasars and the matched sample of typical quasars from T20 along with their respective median 1$\sigma$ error bars. Both quasar samples have been down-sampled using the method outlined in Section~4.1 of T20. Panel (d): The same as panel~(c), however we depict the mini-BAL quasars. \label{fig:1}}
\end{center}
\end{figure}

The primary objective of this work was to investigate the \xray\ variability properties of BAL and mini-BAL quasars on long timescales ($\Delta t \geq 2\times10^6$ rest-frame seconds). Our sample is well suited for investigating such variability since observations over the $\approx20$ year lifetime of \emph{Chandra} were included. After removing the short-timescale epoch permutations and radio-loud quasars, 44 variability amplitudes (37 \xray\ detected in both epochs) of 24 BAL quasars and 48 variability amplitudes (44 \xray\ detected in both epochs) of 35 mini-BAL quasars remained. We also constructed a matched sample of 149 observations of 91 typical quasars from T20 that have statistically similar redshift, 2500~\AA\ luminosity, and timescale distributions to the BAL and mini-BAL quasars (see Figure \ref{fig:A3}). Panel~(a) of Figure~\ref{fig:1} confirms that the BAL quasars in our sample with either 0.5--2 keV or 0.5--7 keV detections are generally \xray\ weaker than both the typical and mini-BAL quasars as expected (e.g.,\ \citealt{Gibson2009}), indicating that our samples are not biased toward \xray\ bright objects. The distribution of log(count-flux ratio) for each of the three distributions is depicted in panel~(b) of Figure \ref{fig:1}, and panels~(c) and (d) depict the log(count-flux ratio) as a function of rest-frame timescale, $\Delta t$, for the BAL quasars and mini-BAL quasars, respectively. Panels (b)--(d) demonstrate that the distributions of \xray\ variability amplitudes of these three quasar populations span a similar range.

To determine the number of BAL and mini-BAL quasars that exhibit significant \xray\ variability beyond that expected from the measurement uncertainties, we computed the $X^2$ statistic as defined by Equation~3 of \citet{Yang2016}. This statistic is similar to the $\chi^2$ statistic, where the observed values are the measured count fluxes in each epoch of the quasar LC, the expected value is the average count flux of the LC, and the measurement uncertainty of each observation is computed as in T20. The number of total counts per-epoch in each \xray\ LC in our sample, however, is not large enough to assume a $\chi^2$ distribution to test the null hypothesis that the \xray\ fluxes do not vary. Instead, to estimate the $p$-value of the null hypothesis, we implemented a Monte Carlo approach (\citealt{Paolillo2004, Young2012}) in which we simulated count fluxes for the epochs in each LC assuming that they follow a Poisson distribution centered on the average flux of the LC, and given the measured effective area and exposure time of the observation. Source extraction was performed on these simulated data, and $X^2$ values were computed for the simulated LCs. This was repeated $10,000$ times to generate a simulated $X^2$ distribution for each quasar LC. The $p$-values of the observed $X^2$ were then determined from these simulated distributions. We found that 8/24 BAL quasars and 12/35 mini-BAL quasars exhibited significant \xray\ variability at the 95\% confidence level. At the same confidence level, \citet{Saez2012} found that only 3/11 BAL quasars varied significantly in the \xray. 

The \xray\ variability amplitude distributions for the BAL and mini-BAL quasars were tested to assess their consistency with a Gaussian distribution. Both distributions were found to be consistent with a Gaussian distribution according to a Shapiro-Wilk test, suggesting that the long-term \xray\ variability of these samples is consistent with random fluctuations rather than being driven by an additional physical process as was found for the typical quasars (see Section~4.2 of T20 for more details). The observed variability distributions for the three populations in panel~(b) of Figure \ref{fig:1}, however, depict the intrinsic distributions broadened by the measurement errors. To compare the \xray\ variability of the three populations properly, we estimated the intrinsic dispersion of the parent population using the method from \citet{Maccacaro1988}. We found that the intrinsic dispersions of the BAL, mini-BAL, and typical quasars are $0.179\pm0.0211$, $0.169\pm0.020$, and $0.158\pm0.014$, respectively. All three intrinsic \xray\ variability distributions are statistically consistent within their 1$\sigma$ errors. 

Since BAL quasars are \xray\ weak due to internal absorption, our a priori thinking was that they might vary more in \hbox{X-rays} than typical quasars if the properties of the absorbing gas often change between epochs (which seems more likely to occur on longer timescales), making our contrary result notable (see \citealt{Saez2012} for further discussion). A larger sample of BAL and mini-BAL quasars with multiple, sensitive \xray\ observations is required to reduce the measurement uncertainties to constrain better their \xray\ variability properties.

\acknowledgments

\appendix

\section{Catalog Production and Supplementary Figures}

Here we describe in more detail the methods used to select BAL and mini-BAL quasars from the Sloan Digital Sky Survey (SDSS; \citealt{York2000})  Data Release 16 quasar catalog (DR16Q; \citealt{Lyke2020}) as well as the \xray\ data processing and source-extraction methods.

We compiled our list of BAL quasars from the DR16Q catalog. This catalog combined all of the spectroscopically confirmed quasars from previous SDSS data releases with new observations to generate a large catalog of $740,414$ bona fide quasars. Included in this catalog is a measurement of the balnicity index (BI; \citealt{Weymann1991}) and the absorption index (AI; \citealt{Hall2002}) which are common measures of the strength of the absorption trough near the \ion{C}{4} emission line. There is no canonical definition separating BAL and mini-BAL quasars in terms of their absorption strength; therefore, we adopt the definition in \citet{Gibson2009} where BAL quasars have BI$>0$ and mini-BAL quasars have BI$=0$ and AI$>0$. Imposing these restrictions, along with a brightness cutoff at $i < 20$, we find a total of 37,207 BAL and mini-BAL quasars. Finally, we searched for counterparts in \emph{Chandra} observations that were observed multiple times using a similar approach to that outlined in Section 2 of \citet{Timlin2020b}, where we restrict the observations to be longer than 5 ks and the quasar to lie less than nine arcmin from the \emph{Chandra} aim-point. We found 293 observations (251 of which were serendipitous, indicating that our sample is not biased substantially by targeted observations; Left panel of Figure \ref{fig:A1}) of 93 BAL and mini-BAL quasars lie within the \emph{Chandra} footprint that have multiple sensitive observations (see Figure \ref{fig:A1}, right panel).

All 293 observations were processed in the same manner as in T20 using CIAO tools \citep{Fruscione2006}. After reprocessing and deflaring, sources were extracted using {\tt{wavdetect}} in soft (0.5--2 keV), hard (2--7 keV), and full (0.5--7 keV) bands. If the \xray\ positions from source extraction matched within 2\arcsec\ of the optically determined position, the \xray\ position was used, otherwise the positions in the DR16 quasar catalog were adopted. Counts were extracted following the method in Section~2 of \citet{Timlin2020a} by centering circular regions on the adopted source position. The radius of these circles increased as the off-axis angle of the source on the detector increased. Background counts were also extracted using circular annuli surrounding the source position. In cases where the background region overlapped with a chip boundary, `pie'-shaped regions were adopted (e.g.,\ \citealt{Pu2020}). Background regions were visually inspected to ensure that no sources were present in the background aperture. Finally, exposure maps were created using the {\tt{flux\_image}} tool as in \citet{Timlin2020b} to quantify the loss of effective exposure for sources at large off-axis angles as well as the degradation of the ACIS quantum efficiency with time and have units of photons$^{-1}$ cm$^2$ s. These exposure maps were created in the soft, hard, and full energy bands (with effective energies of 1, 3, and 2 keV, respectively), and were used to compute the effective exposure times in each band.

As in \citet{Timlin2020b}, we computed the count flux (cts cm$^{-2}$  s$^{-1}$) instead of physical flux (erg cm$^{-2}$  s$^{-1}$) for each observation to mitigate additional uncertainty in the flux that comes from fitting a spectral model (see Section 4.1 of \citealt{Timlin2020b} for more details). The count fluxes were computed in the observed-frame full band (\hbox{0.5--7 keV}). The BAL quasars in this sample span a moderate range in redshift ($z=$ 1.57--3.2) and thus the observed-frame bandpass used in this investigation to compute the count fluxes spans similar rest-frame energies for all the quasars in the sample. Variability amplitudes are computed between every epoch of observation by taking the ratio of the earlier observation to the later observation. Therefore, for each \xray\ light curve with $N$ epochs, there are $N(N-1)/2$ unique permutations of variability amplitudes. To ensure that any single quasar does not dominate the sample due to its large number of epochs, we down-sampled the \xray\ light curves to only 3 observations: the first epoch in the light curve, the last epoch in the light curve, and a randomly drawn epoch between the two (see \citealt{Timlin2020b} for more details). For the final sample discussed in the work, we also remove quasars that are radio loud, where the radio-loudness parameter, $R$, is defined in the same way as in Section~3.2 of \citet{Timlin2020b}. Radio-loud quasars are included and flagged in the catalog presented below.

This catalog contains all 293 observation where each row represents a single observation. The columns presented in the Table are outlined below. Unique quasars can be distinguished through the NAME column, and radio-loud quasars can be removed using the RL\_FLAG. See \citet{Timlin2020b} for more information about the columns.

\begin{itemize}
\item[--] Column (1): SDSS name
\item[--] Column (2): J2000 RA
\item[--] Column (3): J2000 DEC
\item[--] Column (4): Redshift (DR16Q)
\item[--] Column (5): log Galactic column density (cm$^{-2}$; \citealt{Kalberla2005}) 
\item[--] Column (6): Observation ID
\item[--] Column (7): Off-axis Angle (arcmin)
\item[--] Column (8): Effective exposure time (seconds)
\item[--] Column (9): Detection probability in the soft band
\item[--] Column (10): Detection probability in the hard band
\item[--] Column (11): Detection probability in the full band
\item[--] Column (12): Raw counts (soft band)
\item[--] Column (13): Raw background counts (soft band)
\item[--] Column (14): Net counts (soft band)
\item[--] Column (15): $1\sigma$ upper limit on net counts (soft band)
\item[--] Column (16): $1\sigma$ lower limit on net counts (soft band)
\item[--] Column (17): Raw counts (hard band)
\item[--] Column (18): Raw background counts (hard band)
\item[--] Column (19): Net counts (hard band)
\item[--] Column (20): $1\sigma$ upper limit on net counts (hard band)
\item[--] Column (21): $1\sigma$ lower limit on net counts (hard band)
\item[--] Column (22): Raw counts (full band)
\item[--] Column (23): Raw background counts (full band)
\item[--] Column (24): Net counts (full band)
\item[--] Column (25): $1\sigma$ upper limit on net counts (full band)
\item[--] Column (26): $1\sigma$ lower limit on net counts (full band)
\item[--] Column (27): Mean soft-band exposure map pixel value of the source region (photons$^{-1}$ cm$^2$ s)
\item[--] Column (28): Mean soft-band exposure map pixel value of the background region (photons$^{-1}$ cm$^2$ s)
\item[--] Column (29): Mean hard-band exposure map pixel value of the source region (photons$^{-1}$ cm$^2$ s)
\item[--] Column (30): Mean hard-band exposure map pixel value of the background region (photons$^{-1}$ cm$^2$ s)
\item[--] Column (31): Mean full-band exposure map pixel value of the source region (photons$^{-1}$ cm$^2$ s)
\item[--] Column (32): Mean full-band exposure map pixel value of the background region (photons$^{-1}$ cm$^2$ s)
\item[--] Column (33): Chip-edge flag (0 = good detection; 1 = edge detection).
\item[--] Column (34): \emph{Chandra} start time (seconds)
\item[--] Column (35): SDSS Plate ID
\item[--] Column (36): SDSS MJD
\item[--] Column (37): SDSS FIBER ID
\item[--] Column (38): DR16Q BAL probability
\item[--] Column (39): \ion{C}{4} balnicity index
\item[--] Column (40): Uncertainty in the balnicity index
\item[--] Column (41): Absorption index
\item[--] Column (42): Uncertainty in the absorption index
\item[--] Column (43): BAL flag (1=BAL)
\item[--] Column (44): miniBAL flag (1=miniBAL)
\item[--] Column (45): Observed $i$-band magnitude
\item[--] Column (46): Uncertainty on the $i$-band magnitude
\item[--] Column (47): $i$-band extinction
\item[--] Column (48): Absolute $i$-band magnitude (corrected to $z=2$)
\item[--] Column (49):  log 2500 \AA\ monochromatic luminosity (erg s$^{-1}$ Hz$^{-1}$)
\item[--] Column (50): log 2500 \AA\ flux density (erg cm$^{-2}$ s$^{-1}$ Hz$^{-1}$)
\item[--] Column (51): 2 keV flux density (erg s$^{-1}$)
\item[--] Column (52): $1\sigma$ lower limit on the 2 keV flux density (erg cm$^{-2}$ s$^{-1}$ Hz$^{-1}$)
\item[--] Column (53): $1\sigma$ upper limit on the 2 keV flux density (erg cm$^{-2}$ s$^{-1}$ Hz$^{-1}$)
\item[--] Column (54): $\alpha_{\rm{ox}}$ determined using the 2 keV flux density derived from the soft-band 
\item[--] Column (55): $\alpha_{\rm{ox}}$ determined using the 2 keV flux density derived from the full-band
\item[--] Column (56): Targeted observation (1=targeted)
\item[--] Column (57): Radio-loudness flag (1=radio loud, defined as $R>30$; see \citealt{Timlin2020b})
\end{itemize}

\renewcommand{\thefigure}{S1}
\begin{figure}[h!]
\begin{center}
\includegraphics[scale=0.4,angle=0]{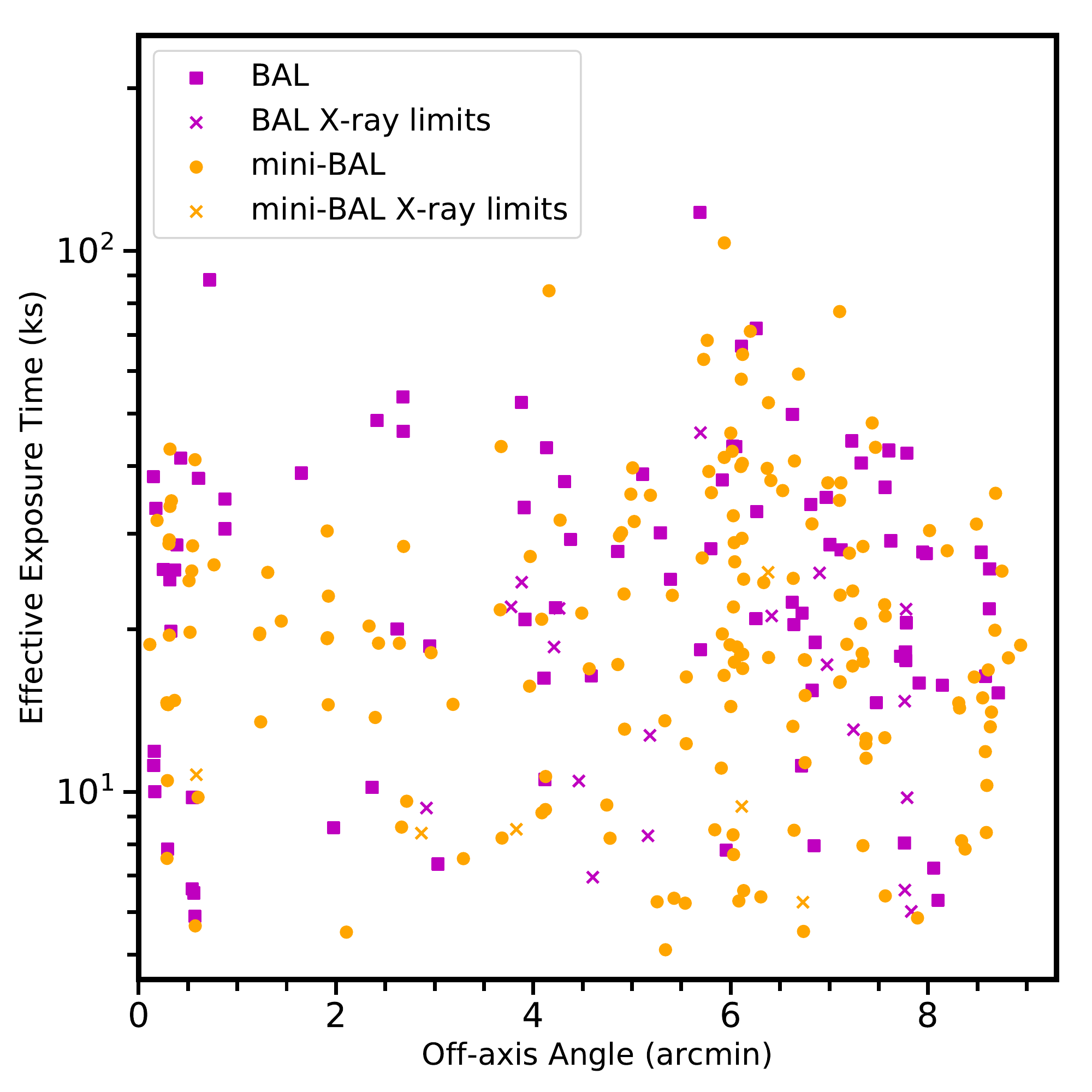}
\includegraphics[scale=0.4,angle=0]{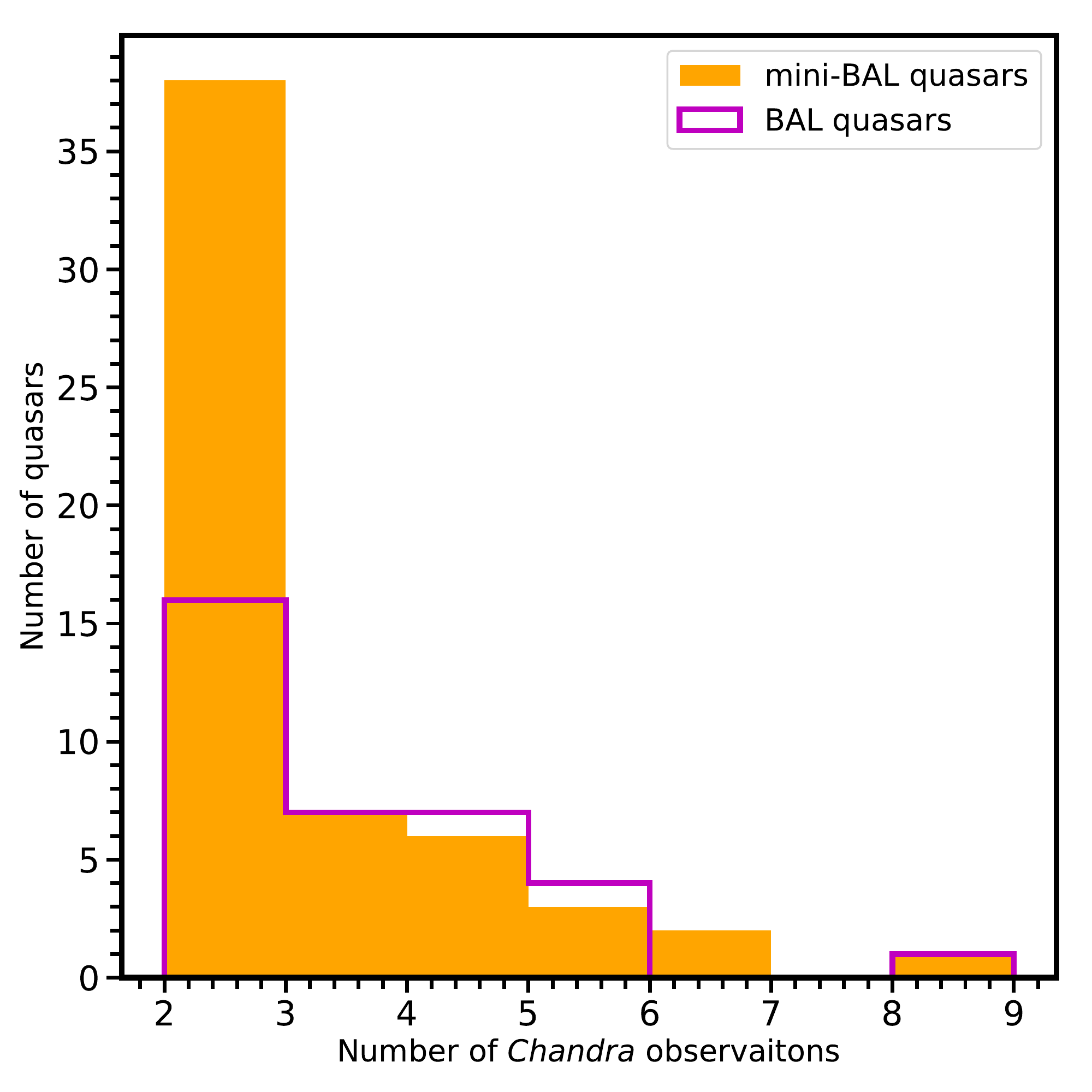}
\caption{Left: Effective exposure time as a function of off-axis angle for the BAL and mini-BAL quasars in the full sample. \xray\ detections of the BAL and mini-BAL quasars are depicted by the magenta squares and orange circles, respectively. Observations in which the quasar was not \xray\ detected are depicted by the crosses of the corresponding color. Right: The number of \emph{Chandra} observations per quasar for the BAL (magenta) and mini-BAL (orange) quasars in our full sample. All 293 observations of the 93 BAL and mini-BAL quasars are reported in the supplementary table. For our final analysis, we down-sampled every quasar LC to contain at most three epochs to mitigate the effects that any one quasar with many observations had on the analysis.  \label{fig:A1}}
\end{center}
\end{figure}

\renewcommand{\thefigure}{S2}
\begin{figure}[h!]
\begin{center}
\includegraphics[scale=0.4,angle=0]{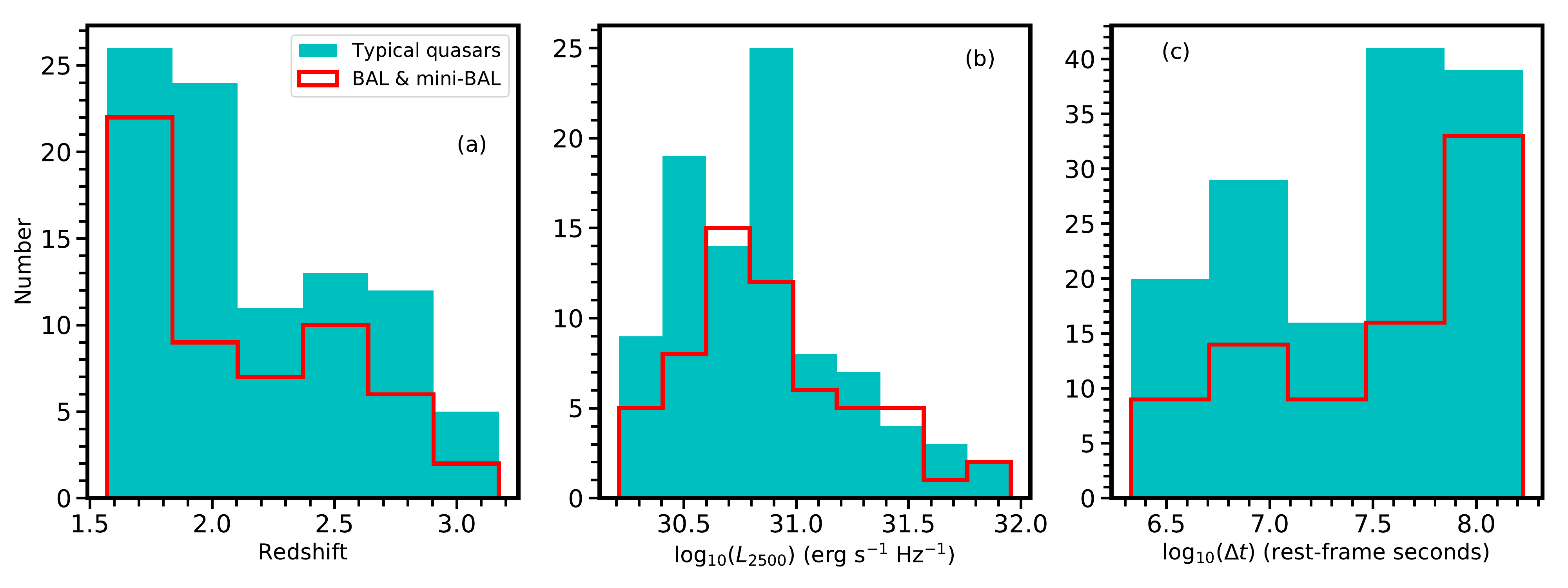}
\caption{Panel (a): Redshift distribution of our sample of BAL and mini-BAL quasars (red open) compared to the matched sample of typical quasars from T20 (cyan filled). An Anderson-Darling test of similarity indicates that these distributions are statistically similar ($p$-value $= 0.907$). Panel (b): Similar to panel~(a), however we demonstrate the similarity of the 2500 \AA\ luminosity between the BAL and mini-BAL quasar sample and the typical quasars ($p$-value $= 0.42$). Panel (c): We depict the timescale distribution of the BAL and mini-BAL quasar sample as well as the timescales of the typical quasars. These distributions are statistically similar ($p$-value $= 0.102$). In all three panels, the quasar LCs have been down-sampled \label{fig:A3}}
\end{center}
\end{figure}

\end{document}